\documentclass[preprint,aps,pra,superscriptaddress,showpacs,floatfix]{revtex4-1}
\usepackage{graphicx}
\usepackage{wrapfig}
\usepackage{dcolumn}
\usepackage{bm}
\usepackage{amsfonts}
\usepackage{amsmath}
\usepackage{cancel}
\usepackage{multirow}
\usepackage{color}
\usepackage{ulem}

\newcommand {\be}{\begin{equation}}
\newcommand {\ee}{\end{equation}}
\newcommand {\bea}{\begin{eqnarray}}
\newcommand {\ea}{\end{eqnarray*}}
\newcommand {\ba}{\begin{eqnarray*}}
\newcommand {\eea}{\end{eqnarray}}

\newcommand {\au} {a.u.}

\begin{document}

\title{Spectra of helium clusters with up to six atoms using soft core potentials}

\author{M. Gattobigio}
\affiliation{Universit\'e de Nice-Sophia Antipolis, Institut Non-Lin\'eaire de
Nice,  CNRS,
1361 route des Lucioles, 06560 Valbonne, France }
\author{A. Kievsky}
\author{M. Viviani}
\affiliation{Istituto Nazionale di Fisica Nucleare, Largo Pontecorvo 3, 56100 Pisa, Italy}

\begin{abstract}
In this work we investigate small clusters of helium atoms using the
hyperspherical harmonic basis. We consider systems with $A=2,3,4,5,6$ atoms
with an inter-particle potential which does not present a strong repulsion at
short distances. We use an attractive gaussian potential that reproduces the
values of the dimer binding energy, the atom-atom scattering length, and the
effective range obtained with one of the widely used He-He interactions, the
LM2M2 potential. In systems with more than two atoms we consider a repulsive 
three-body force that, by construction, reproduces the trimer binding energy
of the LM2M2 potential. With this model, consisting in the sum of
a two- and three-body potential, we have calculated the spectrum of clusters
formed by four, five and six helium atoms. We have found that these systems
present two bound states, one deep and one shallow close to the threshold fixed
by the energy of the $(A-1)$-atom system. Universal relations between the
energies of the excited  state of the $A$-atom system and the ground state
energy of the $(A-1)$-atom system are extracted as well as the ratio between
the ground state of the $A$-atom system and the ground state energy of the
trimer.

\end{abstract}

\pacs{31.15.xj, 31.15.xt, 36.90.+f, 34.10.+x}
\maketitle

\section{Introduction}

Systems composed by few helium atoms have been object of intense investigation
from a theoretical and experimental point of view. The existence of the He-He
molecule was experimentally established using diffraction
experiments~\cite{luo_direct_1996,schoellkopf_nondestructive_1996,schoellkopf_nondestructive_1994,grisenti_determination_2000}.
Its binding energy $E_{2b}$ has been estimated to be around 1 mK and its
scattering length $a_0$ around 190 a.u.  This makes the He-He molecule one of
the biggest diatomic molecules.  On the theoretical side, several He-He
potentials have been proposed; in spite of different details and derivations,
all of them share the common feature of a sharp repulsion below an
inter-particle distance of approximately 5 \au.

Another important characteristic of the He-He interactions is their effective
range $r_0 \approx 13$~\au. Accordingly, the ratio $a_0/r_0$ is large enough
($>10$) to place small helium clusters into the frame of Efimov
physics~\cite{efimov_energy_1970,efimov_weak_1971}.  As shown by Efimov, when
at least two of the two-body subsystems present an infinitely large scattering
length (or zero binding energy) an infinite sequence of bound states (called
Efimov states) appear in the three-body system; their binding energies scale
in a geometrical way and they accumulate at zero energy. The scaling factor,
$\displaystyle\text{e}^{-2\pi/s_0} \approx 1/515.03$, is universal and depends only
on the ratio between particle masses (for three identical bosons $s_0\approx
1.00624$), not on the details of the two-body interaction (see
Ref.~\cite{braaten_universality_2006} for a review).  For a finite $a_0/r_0$
ratio, the number of the Efimov states has been estimated to be
$N=(s_0/\pi){\rm ln}|a_0/r_0|$~\cite{efimov_weak_1971}; for example, the
(bosonic) three $^4$He system presents an excited state just below the
atom-dimer threshold that has been identified as an Efimov state.

Triggered by this interesting fact, several investigations of the helium trimer
have been produced, establishing that its excited state is indeed an
Efimov-like state (see for example
Refs.~\cite{esry_adiabatic_1996,barletta_variational_2001,nielsen_structure_1998}).
In addition, analysis of the atom-dimer collision in the ultracold regime have
also been performed
~\cite{kolganova_ultracold_2009,motovilov_binding_2001,suno_adiabatic_2008}.

One of the main difficulties in solving the quantum mechanical problem in the
case of three helium atoms results to be the treatment of the strong repulsion
at short distances of the He-He potential.  Specific algorithms have been
developed so far to solve this problem. The Faddeev equation has been
opportunely modified~\cite{kolganova_three-body_1998}. Moreover, the
hyperspherical adiabatic (HA) expansion has been extensively used in this case~\cite{suno_adiabatic_2008}
(for a review see Ref.~\cite{nielsen_three-body_2001}). However, due to the
difficulties in treating the strong repulsion, few calculations exist for
systems with more than three helium atoms. For example, in
Ref.~\cite{lewerenz_structure_1997} the diffusion Monte Carlo method has been used
to describe the ground state of He molecules up to 10 atoms, and in Ref.~\cite{blume_monte_2000} 
a Monte Carlo technique has been used to construct the lowest 
adiabatic potential in systems with 3 and up to 10 helium atoms. On the other hand,
description of few-atoms systems using soft-core potentials are currently
operated (see for example Ref.~\cite{von_stecher_correlated_2009}).  

Therefore, the  equivalence between hard- or soft-core-potential descriptions
needs some clarification. In a recent work~\cite{kievsky_helium_2011}, an
attractive He-He gaussian potential has been used to investigate the three
$^4$He system. In absence of direct experimental data, the two-body 
potential has been designed to reproduce the helium
dimer binding energy $E_{2b}$, the He-He scattering length $a_0$, and the
effective range $r_0$ of the LM2M2 potential \cite{aziz_examination_1991}.
This gaussian potential can be considered as a regularized-two-body contact term
in an Effective Field Theory (EFT) approximation of the physics driven by the
LM2M2 potential~\cite{lepage_how_1997}. It should be noticed that
two potentials predicting similar values of $a_0$ and $r_0$ predict similar phase shifts
in the low energy limit and, therefore, even if
their shape is completely different, they describe in an equivalent way the
physical processes in that limit~\cite{lepage_how_1997}. The equivalence is
lost as the energy is increased, when the details of the potential become more
and more important.

Extending the study to the three-body system, differences between the
attractive gaussian and the LM2M2 potentials are immediately observed.  For
example the trimer ground state  energies differ by more than $15\%$ (see
Table~\ref{tab:table2}). A natural way to restore the equivalence between the
two potentials is by the addition of a three-body soft-term force
to the gaussian potential. On the other hand in an
EFT treatment of the three-boson system with large scattering length, a
three-body-contact term is required at leading order (LO). Its strength
is usually determined by fixing a three-body observable as for
example one of the trimer bound state energies. After this choice
cut-off independent results are obtained~\cite{bedaque_renormalization_1999}. Following
this ideas, and based on Ref.~\cite{kievsky_helium_2011}, in the
present work we have considered a gaussian-hypercentral three-body force
with the strength fixed to reproduce the LM2M2 ground state binding energy of
the three-atom system. The quality of this
description has been studied for different ranges of the three-body force.

Using the two-atom and three-atom systems to fix the model interaction,
we have analyzed heavier systems, up to $A=6$ atoms.  The numerical
calculations are performed by means of the hyperspherical harmonic (HH)
expansion method with the technique developed recently by the authors in
Ref.~\cite{gattobigio_nonsymmetrized_2011}. In this approach, the authors 
use the HH basis, without a previous symmetrization procedure, to describe
bound states in systems up to six particles. The method is based in a
particular representation of the Hamiltonian matrix, as a sum of products of
sparse matrices, well suited for a numerical implementation. Converged results
for different eigenvalues, with the corresponding eigenvectors belonging to
different symmetries, have been obtained. As a novelty, in the present work we
extend the formalism to treat a three-body force.
Moreover, as we are dealing with atoms of $^4$He, only the
the spectrum corresponding to totally symmetric eigenstates are of interest. 

After fixing the strength of the three-body force to correctly describe the
LM2M2 three-body ground state $E_{3b}^{(0)}=126.4$ mK, we have calculated the
first three levels of the spectrum with total angular momentum $L=0$ of the
$A=4,5,6$ systems. In the three cases we have found that the first two levels
are bosonic bound states, one deep, $E^{(0)}_{Ab}$, and one very shallow,
$E^{(1)}_{Ab}$, close to the threshold formed by the $A-1$ system plus one atom.
The third state in all cases belongs to a mixed symmetry with an energy above
the threshold and therefore not representing a bound state.  The appearance of
only two bound states in this systems is in agreement to previous
calculations~\cite{blume_monte_2000}.  This fact has been observed in $A=4$ and
interpreted as a consequence of the Efimov like spectrum of the $A=3$
system~\cite{hammer_universal_2007}.  It should be noticed that, whereas
converged results can be found in the literature for the ground state of the
many atom systems, the energy of the excited states is much more difficult to
calculate and only rough estimates are available. 

To gain insight on the shallow state, we have varied the range of the
three-body force (maintaining fixed the three-body ground state energy) and we
have studied the effects of that variation in the $A=4,5,6$ spectrum. In the
range considered, the variation produces small effects in the eigenvalues;
however, it is crucial to determine if the shallow state is bound or not with
respect to the $A-1$ threshold.  Interestingly, we have observed that when the
ranges of the two- and three-body forces have a ratio of about $\sqrt{2}$, the
ratio between the shallow- and ground-state energy is
$E^{(1)}_{Ab}/E^{(0)}_{A-1b}\approx 1.01-1.02$, in agreement with
Refs.~\cite{von_stecher_signatures_2009,deltuva_efimov_2010}. This analysis
confirms previous observations that each Efimov state in the $A=3$ system
produces two bound states in the $A=4$ system. 
Furthermore, we have found $E_{4b}^{(0)}/E_{3b}^{(0)}\approx 4.5$,
$E_{5b}^{(0)}/E_{3b}^{(0)}\approx 10.5$, and 
$E_{6b}^{(0)}/E_{3b}^{(0)}\approx 18.5$, which is in 
agreement with Refs.~\cite{deltuva_efimov_2010,von_stecher_weakly_2010}.

The paper is organized as follows. In the next Section II we describe the two-
and three-body forces we used in our calculations to reproduce LM2M2 data.  In
Section III the results for the bound states of the $A=3,4,5,6$ He clusters are
collected whereas the conclusions are given in the last section.
Some technical details of the method have been summarized in the Appendix.

\section{Soft-core two- and three-body helium potential}

As mentioned in the Introduction, the $^4$He-$^4$He interaction presents a
strong repulsion at short distances, below 5 a.u. This characteristic makes
it difficult a detailed description of the system with more than four atoms.
Accordingly, in the present work we have studied small clusters of helium
interacting through a soft-core two- and three-body potentials which can be 
interpreted as regularized two- and three-body contact terms in a LO-EFT 
approximation of LM2M2.

Following Refs.~\cite{nielsen_structure_1998,kievsky_helium_2011} we use the gaussian
two-body potential
\begin{equation}
V(r)=V_0 \,\, {\rm e}^{-r^2/R^2}\,,
\label{twobp}
\end{equation}
with $V_0=-1.227$ K and $R=10.03$~a.u.. In the following we use
$\hbar^2/m=43.281307~\text{(a.u.)}^2\text{K}$. This parametrization of the
two-body potential approximately reproduces the dimer binding energy $E_{2b}$,
the atom-atom scattering length $a_0$ and the effective range $r_0$ given by
the LM2M2 potential.  Specifically, the results for the gaussian potential are
$E_{2b}=-1.296$ mK, $a_0=189.95$ a.u. and $r_0=13.85$ a.u., to be compared to
the corresponding LM2M2 values $E_{2b}=-1.302$ mK, $a_0=189.05$ a.u. and
$r_0=13.84$ a.u..  As shown in Ref.~\cite{kievsky_helium_2011}, the use of the
gaussian potential in the three-atom system produces a ground state binding
energy $E^{(0)}_{3b}=150.4$ mK, which is appreciable bigger than the LM2M2 helium
trimer ground state binding energy of $126.4$ mK.  A smaller difference, though
still appreciable, is observed in the first excited state (see
Table~\ref{tab:table2}). 

In order to have a closer description to the $A=3$ system obtained with the
LM2M2 potential, we introduce the following three-body interaction
\begin{equation}
W(\rho_{ijk})=W_0 \,\, {\rm e}^{-2\rho^2_{ijk}/\rho^2_0}\,,
  \label{eq:hyptbf}
\end{equation}
where $\rho^2_{ijk}=\frac{2}{3}(r^2_{ij}+r^2_{jk}+r^2_{ki})$ is the three-body
hyperradius in terms of the distances of the three interacting particles.
Moreover, the strength $W_0$ is fixed to reproduce the LM2M2 helium trimer binding
energy of $126.4$ mK.  We have studied different cases by varying the range of
the three-body force $\rho_0$ between 4 and 16 a.u.. Specific cases with the
corresponding results in the $A=3$ system are shown in Table~\ref{tab:table2}.
In the first two rows of the table we report the ground- and
excited-binding energies of the trimer, both for  LM2M2 potential and its
gaussian representation. The excess of binding is evident for this
last model. Successively, we report, for selected values of $W_0$ and $\rho_0$,
the binding energies obtained summing to the gaussian potential the (repulsive)
three-body force. By construction the ground state has been fixed to the LM2M2
value and, in addition, we can observe that the excited state $E^{(1)}_{3b}$ is now
closer to the corresponding LM2M2 result, showing an extremely small variation
with $\rho_0$; the difference between the extremal values obtained for 
$\rho_0=4~\text{a.u.}$ and  $\rho_0=16~\text{a.u.}$ is less than 1\%.

It should be noticed that the ranges $R$ of the two-body force and $\rho_0$ of
the three-body force are somehow related.  The gaussian two-body force can be
thought as originating from a contact interaction regularized using a gaussian
cutoff with $\Lambda=R^{-1}$ (see for example
Ref.~\cite{braaten_universality_2006}). This means that configurations, in the
$A=2$ system, in which two atoms have a relative momentum $k>\Lambda$ remain
outside the present description or, in other words, details of the interaction
for distances below $\sqrt{\langle r^2 \rangle} = R/\sqrt 2$ are not accessible. In the $A>2$ systems, three atoms
interact through the three-body force when they happen to be inside a sphere
of radius $\rho_0/\sqrt 2$ at the same time. It is clear that, as no information is
introduced in the two-body system for distances below $R/\sqrt 2$, from the relation
$\rho^2_{ijk}=\frac{2}{3}(r^2_{ij}+r^2_{jk}+r^2_{ki})$ and putting each
distance at the value $R/\sqrt 2$, we obtain for the three-body range $\rho^2= R^2$.
Since $R$ has been fixed in order to describe two-body quantities, in the
description of systems with $A>3$ atoms we consider different values of $\rho_0$ with
particular attention at the region $\rho_0\approx \sqrt{2}R$.

The calculations for the $A\ge 3$ systems, up to six atoms, are performed using
the unsymmetrized HH basis. The method has been recently used to describe up to
six nucleons interacting through a central
potential~\cite{gattobigio_nonsymmetrized_2011}. A brief description of the
method is given in the Appendix. The novelty presented here regards the
implementation of the three-body force. Using a particular rotation of the HH
basis it is possible to construct the potential energy as a product of sparse
matrices. The Hamiltonian matrix is obtained using the following orthonormal
basis
\begin{equation}
  \langle\rho\,\Omega\,|\,m\,[K]\rangle =
  \bigg(\beta^{(\alpha+1)/2}\sqrt{\frac{m!}{(\alpha+m)!}}\,
  L^{(\alpha)}_m(\beta\rho)
  \,{\text e}^{-\beta\rho/2}\bigg)
  {\cal Y}^{LM}_{[K]}(\Omega_N)  \,,
  \label{mhbasis}
\end{equation}
where $L^{(\alpha)}_m(\beta\rho)$ is a Laguerre polynomial with $\alpha=3N-1$
and $\beta$ a variational non-linear parameter.  The matrix elements of the
Hamiltonian are obtained after integrations in the $\rho,\Omega$ spaces. They
depend on the indices $m,m'$ and $[K],[K']$ as follows
\begin{equation}
\begin{aligned}
  \langle m'\,[K']|H|\,m\,[K] \rangle = -\frac{\hbar^2\beta^2}{m}
 ( T^{(1)}_{m'm}-K(K+3N-2) T^{(2)}_{m'm}) \delta_{[K'][K]} \cr
 + \sum_{i<j} \left[
\sum_{[K''][K''']}{\cal B}^{ij,LM}_{[K][K'']}{\cal B}^{ij,LM}_{[K'''][K']}
 V^{m,m'}_{[K''][K''']}\right] 
 + \sum_{i<j<k} \left[
\sum_{[K''][K''']}{\cal D}^{ijk,LM}_{[K][K'']}{\cal D}^{ijk,LM}_{[K'''][K']}
 W^{m,m'}_{[K''][K''']}\right] \,.
\end{aligned}
\label{eq:hmm}
\end{equation}
The matrices $T^{(1)}$ and $T^{(2)}$ have an analytical form and are given in
Ref.~\cite{gattobigio_harmonic_2009}.  The matrix elements $V^{m,m'}_{[K][K']}$
and $W^{m,m'}_{[K][K']}$ are obtained after integrating the matrices
$V_{12}(\rho)$ and $W(\rho)$ in $\rho$-space (we will call the corresponding
matrices $V_{12}$ and $W$).  Introducing the diagonal matrix $D$ such that
$\langle [K']\,|\,D\, | [K]\rangle = \delta_{[K],[K']} K(K+3N-2)$, and the
identity matrix $I$ in $K$-space, we can rewrite the Hamiltonian schematically
as
\begin{equation}
  H = -\frac{\hbar^2\beta^2}{m} ({}^{(1)}T \otimes I  +  {}^{(2)}T\otimes D )
  + \sum_{i<j} [{\cal B}^{LM}_{ij}]^t\, V_{12}\,{\cal B}^{LM}_{ij} \,,
  + \sum_{i<j<k} [{\cal D}^{LM}_{ijk}]^t\, W\,{\cal D}^{LM}_{ijk} \,,
  \label{eq:schemH}
\end{equation}
in which the tensor product character of the kinetic energy is explicitly
given. A method to diagonalize a matrix of this form is given in
Ref.~\cite{gattobigio_nonsymmetrized_2011}.

\section{Results for $A=4,5,6$ He clusters}

In this sections we present results for small clusters, up to $A=6$, formed by
atoms of $^4$He.  Despite the differences observed at the level of the
three-body system between the gaussian two-body force and the LM2M2 potential,
it is of interest the computation of the spectrum produced by the gaussian
two-body force only for the $A=4,5,6$ systems. 

In Table~\ref{tab:table1} we show the $L=0$ ground state $E^{(0)}_{Ab}$ and the
first two excited states $E^{(1)}_{Ab}$ and $E^{(2)}_{Ab}$ for increasing
values of the grand-angular momentum $K$ using the unsymmetrized HH basis.  The
calculations have been performed up to $K=40$ in $A=4$, $K=24$ in $A=5$ and
$K=22$ in $A=6$. It is a property of the HH basis that when all states having a
fixed value of $K$ are included in the expansion of the wave function, the
symmetry of the eigenvectors reflects the symmetries present in the
Hamiltonian.  Since the Hamiltonian is symmetric under exchange of the
particles, the obtained eigenvectors have well defined particle permutation
symmetry. In the present case the ground state $E^{(0)}_{Ab}$ and first excited state
$E^{(1)}_{Ab}$ of the
Hamiltonian matrix for $A=4,5,6$ are symmetric states and belong to the
irreducible representation $[\bm \lambda]$ with $\bm \lambda= \bm A$. In all
cases the second excited state $E^{(2)}_{Ab}$ has mixed symmetry and belongs to the
irreducible representation $[\bm \lambda\,\, \bm{1}]$ with $\bm \lambda= \bm
{A-1}$.  In Table~\ref{tab:table1}  we also observe that the ground state
binding energy, $E^{(0)}_{Ab}$, has a very fast convergence in terms of $K$ and
can be determined with five digits; this value fixes the threshold of the
continuum spectrum in the $A+1$ system.  True bound states in the $A=4$ systems
are those having a binding energy bigger than the trimer binding energy of
$150.4$ mK and, looking at the table, bound states in the $A=5,6$ systems
appears below the threshold of $751.38$ mK and $1945.2$ mK respectively. Since
in all cases the second excited state $E^{(2)}_{Ab}$ results to be above the
threshold, only two bosonic states are bound in the $A=4,5,6$ systems, one deep
and one shallow close to the $A-1$ threshold. The next bosonic state appears
above $E^{(2)}_{Ab}$ and, therefore, it is not bound. This result confirms
previous analysis in the four body sector that the lower Efimov state in the
$A=3$ system produces two bound states, one deep and one shallow.  Here, we
have extended this observation up to the $A=6$ system.  The convergence of the
$E^{(1)}_{Ab}$ is much slower than for the ground state, however with the
extended based used it has been determined with an accuracy well below $1\%$.

For this atom-atom potential the ratio $r_0/a\approx 1/14$ and therefore we are
not too far from the unitary limit, and we can make predictions for the
universal ratios $E^{(1)}_{Ab}/E^{(0)}_{(A-1)b}$ and
$E^{(0)}_{Ab}/E^{(0)}_{3b}$. From the table we can observe that
$E^{(1)}_{4b}/E^{(0)}_{3b}=1.085$, $E^{(1)}_{5b}/E^{(0)}_{4b}=1.10$ and
$E^{(1)}_{6b}/E^{(0)}_{5b}=1.13$. These results are not so close to the
universal ratio of around $1.01$ indicating that effective range corrections
are important. For the ratios with respect to the trimer ground state we have,
$E^{(0)}_{4b}/E^{(0)}_{3b}=5.01$, $E^{(0)}_{5b}/E^{(0)}_{3b}=12.97$ and
$E^{(0)}_{6b}/E^{(0)}_{3b}=25.4$. As we will see, these ratios are
substantially modified when a three-body force is included.

Now, we consider the model with both two- and three-body interaction.  The
pattern of convergence for the ground and excited states of the $A=4,5,6$
helium systems, using the gaussian two-body potential plus the repulsive
three-body potential with $\rho_0=14$ a.u., is given in Table~\ref{tab:table3}.
The maximum grand angular momentum considered is $K=40$ for $A=4$, $K=24$ for
$A=5$ and $K=22$ for $A=6$.  As in the case in which only the two-body force
has been considered, in all of the three cases only two bound states appears,
one deep and one shallow very close to the $A-1$ threshold. The ground state
presents a fast convergence with $K$ and the accuracy can be estimate below
$0.1$~mK. The convergence for the excited state is slower and, for the values
of $K$ considered, its accuracy is given at the level of $3$~mK. However, from
the results it is well established that, with the value of $\rho_0$ considered,
the excited state, $E^{(1)}_{Ab}$, is bound with respect to the $A-1$
threshold.  In fact, for $A=4$ the $3+1$ threshold appears at 126.4~mK and the
upper bound estimate for this state is $129$ mK. Its ratio
$E^{(1)}_{4b}/E^{(0)}_{3b}$ is 1.020.  For $A=5$ the $4+1$ threshold appears at
568.8 mK and the upper bound estimate for the excited state is $575$ mK. Its
ratio $E^{(1)}_{5b}/E^{(0)}_{4b}$ is 1.011.  For $A=6$ the $5+1$ threshold
appears at $1326.6$ mK and the upper bound estimate for the excited state is
$1350$ mK.  Its ratio $E^{(1)}_{6b}/E^{(0)}_{5b}$ is 1.018.  The ratio between
the trimer ground state and the ground states of the $A=4,5,6$ systems are
$E^{(0)}_{4b}/E^{(0)}_{3b}=4.5$, $E^{(0)}_{5b}/E^{(0)}_{3b}= 10.5$ and
$E^{(0)}_{5b}/E^{(0)}_{3b}=18.5$, respectively. These ratios are in good
agreement with those given in
Refs.~\cite{von_stecher_signatures_2009,deltuva_efimov_2010,von_stecher_weakly_2010},
and represent a substantial improvement with respect to the case in which 
the three-body force is not included. At the ratio $r_0/a$ under consideration
the use of the two-body soft-core potential alone reduces the Efimov character
of ground and first-exited states, which is recovered by including the 
three-body force.
It is interesting to compare the results obtained using the soft-core
representation of the LM2M2 potential with the results of
Refs.~\cite{lewerenz_structure_1997,blume_monte_2000} (quoted in
Table~\ref{tab:table3}) obtained using the original LM2M2 interaction. For the
ground state the agreement is around $2\%$ for $A=4,5$ and around $1\%$ for
$A=6$.  The agreement is worst for the excited state, however the results from
Ref.~\cite{blume_monte_2000} are obtained using approximate solutions of the
adiabatic hyperspherical equations. 

The overall agreement for $A=4,5,6$ between LM2M2 and soft
potential gives a further indication that at the LO in an EFT
approach to the Efimov physics there is no need of a four-body force; this is
only a side observation which is not at all conclusive for the lack of
systematic study as a function of the cut-off.

\begin{figure}
  \begin{center}
 \includegraphics[width=\linewidth]{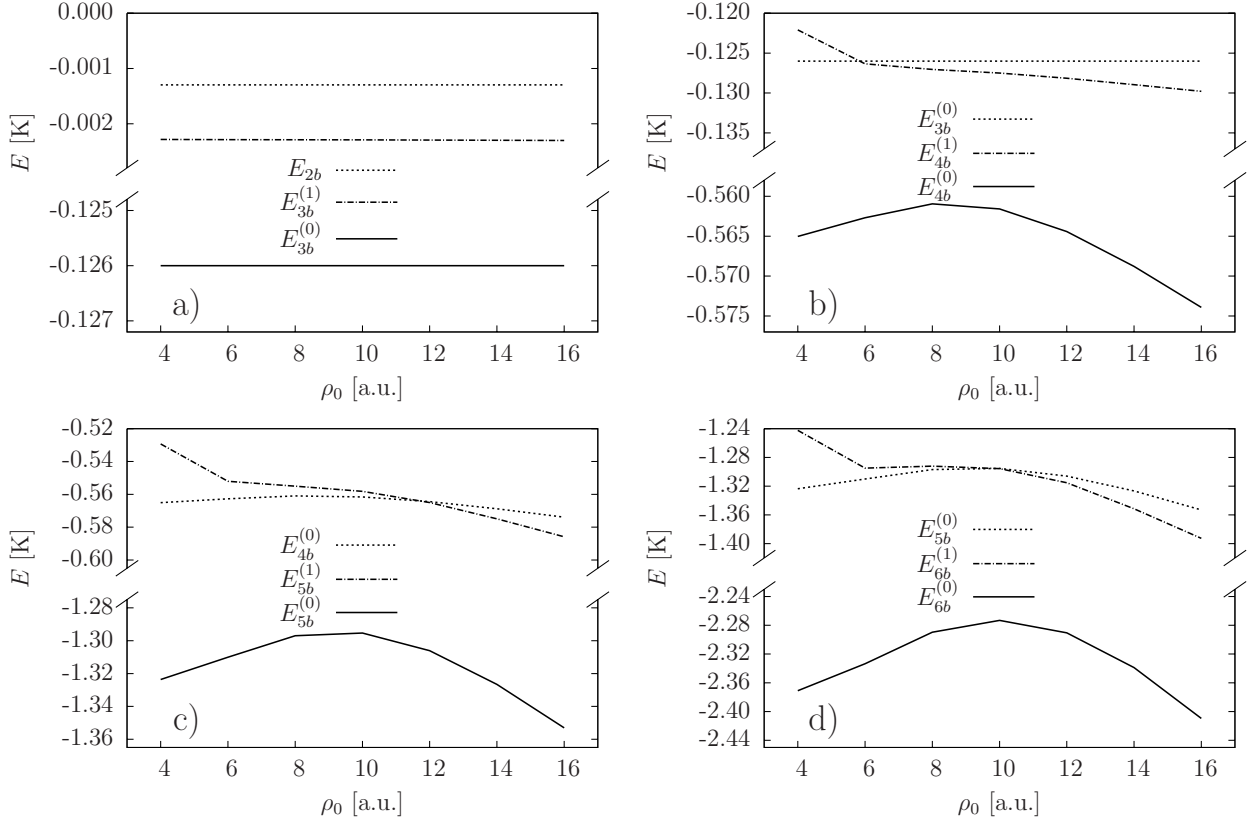}
  \end{center}
  \caption{Ground- and excited-state energies of the $A=2,3,4,5,6$
 systems as a function of $\rho_0$. In panel a) we report the ground- and excited-state
 energy of $A=3$ system together with the ground-state energy of $A=2$; for all of the 
 values of $\rho_0$ we have considered, the excited $A=3$ state is bounded.
 In panel b) we report the ground- and excited-state
 energy of $A=4$ system together with the ground-state energy of $A=3$; the
 excited $A=4$ state is bounded for $\rho_0> 7$.
 In panel c) we report the ground- and excited-state
 energy of $A=5$ system together with the ground-state energy of $A=4$; the
 excited $A=5$ state is bounded for $\rho_0> 12$.
 In panel d) we report the ground- and excited-state
 energy of $A=6$ system together with the ground-state energy of $A=5$; the
 excited $A=6$ state is bounded for $\rho_0> 10$.
 }
 \label{fig:frho}
\end{figure}

Moreover, in the four panels of Fig.~\ref{fig:frho} we analyze modifications to
the spectrum of the systems we have considered when different values of $W_0$
and $\rho_0$ are used. The results for $A=3$ can be extracted from
Table~\ref{tab:table1}; the $A=3$ ground state is stable by construction, and
small variations are observed for $E^{(1)}_{3b}$. As shown in
Fig.~\ref{fig:frho}a, $E^{(1)}_{3b}$ is always below the 2+1 threshold.  For
$A=4$, see Fig.~\ref{fig:frho}b, the excited state $E^{(1)}_{4b}$ is above the
3+1 threshold, and therefore not bounded for values of $\rho_0 < 7$ a.u. For
$A=5$, Fig.~\ref{fig:frho}c, and $A=6$, Fig.~\ref{fig:frho}d, the corresponding
excited states are above the 4+1 and 5+1 thresholds for for values of $\rho_0
<12$~a.u., and $\rho_0<10$~a.u., respectively.  For $A=5,6$ the results for the
bound state present a bump with the smaller binding energy around $\rho_0=10$
a.u..  To sum up, the most sensitive property of the spectrum as a function of   $\rho_0$ is,
for $A=4,5,6$, if the $E^{(1)}_{Ab}$ is above or below threshold. As previously
discussed, a reasonable  choice is $\rho_0=14$~a.u., and around this value all
the excited states are bound.

Finally, in Table~\ref{tab:table4} the results for the universal ratios are
shown in for values of $\rho_0=12,14,16$~a.u.; we observe that small variation
of $\rho_0$ do not drastically change these values. It should be noticed that
in the present analysis the unitary limit is not completely reached since the
ratio between the two-body effective range and scattering length is
$r_0/a\approx 1/14$. An analysis of the universal ratios as a function of $a$
is in progress.

\section{Conclusions}

In this work we have attached two different problems. From one side, we have
studied the possibility of calculating bound and excited states in a bosonic
(atomic) system up to $A=6$ using the unsymmetrized HH expansion and
considering soft two- and three-body forces. On the other hand, the model has
been constructed to approximate the description of small helium clusters taking
as a reference the results of the LM2M2 potential. These two problems are
related since the LM2M2 presents a strong repulsion at short distances.
Therefore, the possibility of using a soft-core representation of the original
potential has been analyzed in detail. In Ref.~\cite{kievsky_helium_2011}
bound states and low energy scattering states of the trimer have been analyzed
using the soft-core representation of the LM2M2. The results obtained in that
work were encouraging in the sense that they were found to be in close
agreement to those obtained using the original potential. 

Here, we have extended the analysis to bigger systems. Therefore, the
description of such systems with sufficient accuracy is of the main importance.
To this end, we have used a method recently developed in
Ref.~\cite{gattobigio_nonsymmetrized_2011} in which the HH basis is used
without symmetrization of the basis states. The basis is complete and, when all
basis elements are included up to a certain maximum value of the grand angular
momentum $K$, the eigenvectors reflect the symmetries present in the
Hamiltonian. In the particular case here considered, the eigenvectors have well
defined symmetry under particle permutation and they can be organized as belonging
to the different irreducible representations of the group of permutations of
$A$ objects, $S_A$. 
This simple fact has allowed to make an important statement regarding the
number of bosonic bound states present in the systems under study. After the
direct diagonalization of the $A$-body system we have analyzed the first three
states for increasing values of $K$. We have considered very extended
basis, up to $K=40$ for the $A=4$ system and $K=24$ ($K=22$) for the
$A=5$ ($A=6$) system. This allowed to obtain converged results
for the first eigenvalues of the spectrum.
The first two were symmetric states having eigenvalues with energy
below the continuum threshold (fixed by the lowest bound state in the $A-1$
system) and therefore they represent true bound states. The third state was
found to belong to a mixed symmetry and results to be above the threshold. This
was the case for all the systems considered ($A=4,5,6$) and it means that the
next bosonic state has an energy above the mixed state and therefore it is not
bound. Therefore we have unambiguously determined that these systems present
only two bound states. 

The two bosonic bound states has been studied for different values of the
three-body potential rage $\rho_0$. This analysis is given
in Fig.~\ref{fig:frho} where the position of the excited state moves
from unbound to bound as $\rho_0$ increases. The particular case
$\rho_0\approx \sqrt2 R$ is explicitly given in Table~\ref{tab:table1}
showing that in fact the excited state is slightly bound. Moreover,
since the He-He potential predicts a large two-body scattering length, we have
studied the universal ratios $E^{(0)}_{Ab}/E^{(0)}_{3b}$ and
$E^{(1)}_{Ab}/E^{(0)}_{(A-1)b}$. 
These ratios have been studied in detail in the $A=4$ case (see
Refs.~\cite{hammer_universal_2007,von_stecher_signatures_2009,deltuva_efimov_2010}).
Estimates have also been obtained for bigger
systems~\cite{von_stecher_weakly_2010}. Our calculations, obtained for one
particular value of the ratio $r_0/a$, are in agreement with those references
on the universal character of these states in $A>3$ systems.  An analysis of
the universal ratios as $a\rightarrow\infty$ is at present under way.

Finally, we would like to discuss the quality of the description using the two-
and three-body soft-core-potential model.  We observe a substantial good
agreement, at the level of 2\% or better, for the ground states of the $A$-atom
systems in comparison to the results of the LM2M2 potential given by Lewerenz
(Ref.~\cite{lewerenz_structure_1997}).  The excited states have been calculated
in Ref.~\cite{blume_monte_2000} though using a reduced Hilbert space. Comparing
to those results we observe an agreement around 5\%.  From this analysis we can
conclude that a four-body force will have effects beyond this level of
accuracy. A deeper analysis in this subject is in progress.
\section{Appendix}

Following Refs.~\cite{gattobigio_nonsymmetrized_2011,gattobigio_harmonic_2009},
we present a brief overview of the properties of the HH basis and its
implementation without generating basis elements with well defined
permutational symmetry.  This approach allows to avoid the complications
of symmetry-adapted-basis construction, and to easily treat permutational-symmetry-breaking 
terms~\cite{gattobigio_nonsymmetrized_2011,gattobigio_fewbody_2011}

We start with the following definition of the Jacobi
coordinates for an equal mass $A$ body system with Cartesian coordinates
$\mathbf r_1 \dots \mathbf r_A$

\begin{equation}
  \mathbf x_{N-j+1} = \sqrt{\frac{2 j}{j+1} } \,
                  (\mathbf r_{j+1} - \mathbf X_j)\,,
   \qquad
   j=1,\dots,N\,.
  \label{eq:jc2}
\end{equation}
For a given set of Jacobi coordinates $\mathbf x_1, \dots, \mathbf x_N$, 
we can introduce the hyperradius $\rho$
\begin{equation}
  \rho = \bigg(\sum_{i=1}^N x_i^2\bigg)^{1/2}
   = \bigg(2\sum_{i=1}^A (\mathbf r_i - \mathbf X)^2\bigg)^{1/2}
   = \bigg(\frac{2}{A}\sum_{j>i}^A (\mathbf r_j - \mathbf r_i)^2\bigg)^{1/2} \,,
  \label{}
\end{equation}
and the hyperangular coordinates $\Omega_N$
\begin{equation}
  \Omega_N = (\hat {\bm x}_1, \dots, \hat {\bm x}_N, \phi_2, \dots, \phi_N) \,,
  \label{}
\end{equation}
with the hyperangles $\phi_i$ defined via
\begin{equation}
  \begin{aligned}
    &x_N = \rho \cos\phi_N \\
    &x_{N-1} = \rho \sin\phi_N \cos\phi_{N-1} \\
    &\qquad\vdots \\
    &x_{i} = \rho \sin\phi_N \cdots \sin\phi_{i+1}\cos\phi_i\\
    &\qquad\vdots \\
    &x_{2} = \rho \sin\phi_N \cdots \sin\phi_{3}\cos\phi_2  \\
    &x_{1} = \rho \sin\phi_N \cdots \sin\phi_{3}\sin\phi_2  \,. \\
  \end{aligned}
  \label{eq:hyp1} 
\end{equation}

\begin{figure}
  \begin{center}
 \includegraphics[width=0.9\linewidth]{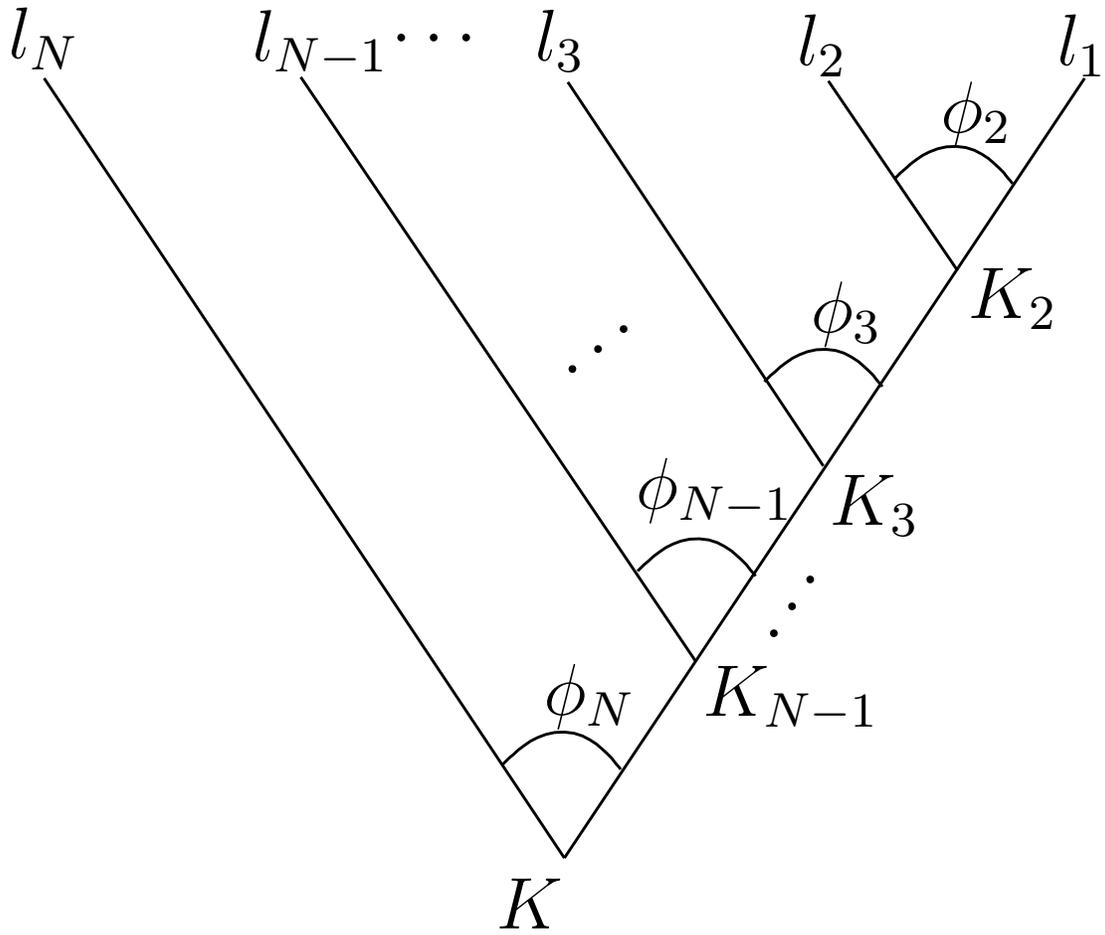}
  \end{center}
  \caption{Hyperspherical tree corresponding to
Eq.({~\protect\ref{eq:hyp1}})}
 \label{fig:tree}
\end{figure}

The explicit expression for the HH functions, 
having well defined values of $LM$, is
\begin{equation}
    {\mathcal Y}^{LM}_{[K]}(\Omega_N) =
    \left[\prod_{j=2}^N 
{\mathcal P}_{K_j}^{\alpha_{l_j},\alpha_{K_{j-1}}}(\phi_j)\right]
    \bigg[Y_{l_1}(\hat {\bm x}_1) \otimes Y_{l_2}(\hat {\bm x}_2)|_{L_2}
    \ldots \otimes Y_{l_{N-1}}(\hat {\bm x}_{N-1})|_{L_{N-1}} 
    \otimes Y_{l_N}(\hat {\bm x}_N) \bigg]_{LM}  \,,
\label{eq:hh0}
\end{equation}
with the indicated coupling scheme. The hyperspherical polynomial is
\begin{equation}
{\mathcal P}_{K_j}^{\alpha_{l_j},\alpha_{K_{j-1}}}(\phi_j) = 
{\mathcal N}_{n_j}^{\alpha_{l_j},\alpha_{K_j}} 
(\cos\phi_j)^{l_j} (\sin\phi_j)^{K_{j-1}} 
P^{\alpha_{K_{j-1}}, \alpha_{l_j}}_{n_j}(\cos2\phi_j) \,.
\end{equation}
The set of
quantum numbers $[K]$ includes the $n_2\ldots n_N$ indices of the Jacobi
polynomials, the $l_1\ldots l_N$ angular momenta of the particles and the
intermediate couplings $L_2 \ldots L_{N-1}$. The $K_j$ quantum numbers are
defined as
\begin{equation}
  K_j = \sum_{i=1}^j (l_i + 2n_i)\,, \qquad n_1 = 0\,, \qquad K \equiv K_N\,,
  \label{}
\end{equation}
$K\equiv K_N$ is known as the grand angular momentum, and ${\cal
N}_{n}^{\alpha\beta}$ is a normalization factor.  For the definition of the
$\alpha_a$, where $a$ can be either an angular momentum $l_j$ or a quantum
number $K_j$, one needs to introduce the hyperspherical-binary-tree
structure~\cite{vilenkin_eigenfunctions_1966}.  For example the tree of
Fig.~\ref{fig:tree} corresponds to the choice of hyperangles given by
Eq.~(\ref{eq:hyp1}), in which the coefficients specializes to $\alpha_{K_j} =
K_j +3j/2-1$ and $\alpha_{l_j}=l_j+1/2$.

Hyperspherical functions constructed using different hyperspherical-coordinate 
definitions can be related using the ${\cal T}$-coefficients~\cite{kildyushov_hyperspherical_1972,kildyushov_n-body_1973}.  
Schematically, these coefficients relate the following tree structures
\begin{equation}
   \begin{minipage}{0.25\linewidth}
  \includegraphics[width=\linewidth]{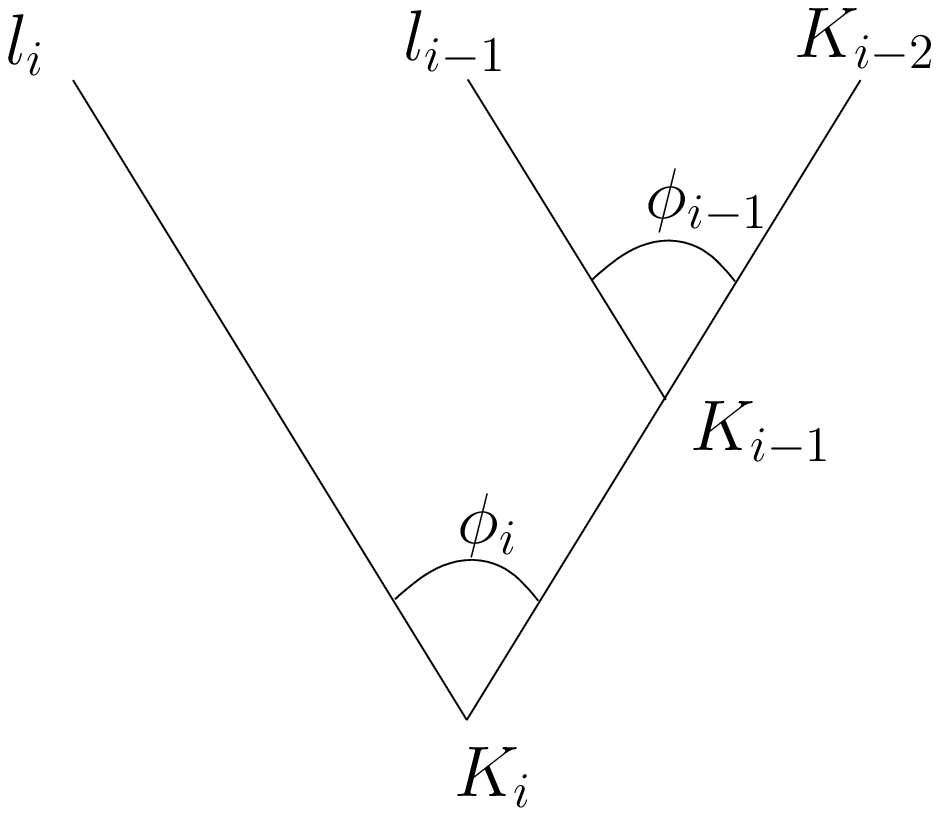}
   \end{minipage}
   =
   \sum_{\tilde n_{i-1}=0}^{N_i}
   {\cal T}^{\alpha_{K_{i-2}}\alpha_{l_{i-1}}\alpha_{l_i}}_{n_{i-1} \tilde n_{i-1}
   K_i}
   \begin{minipage}{0.25\linewidth}
  \includegraphics[width=\linewidth]{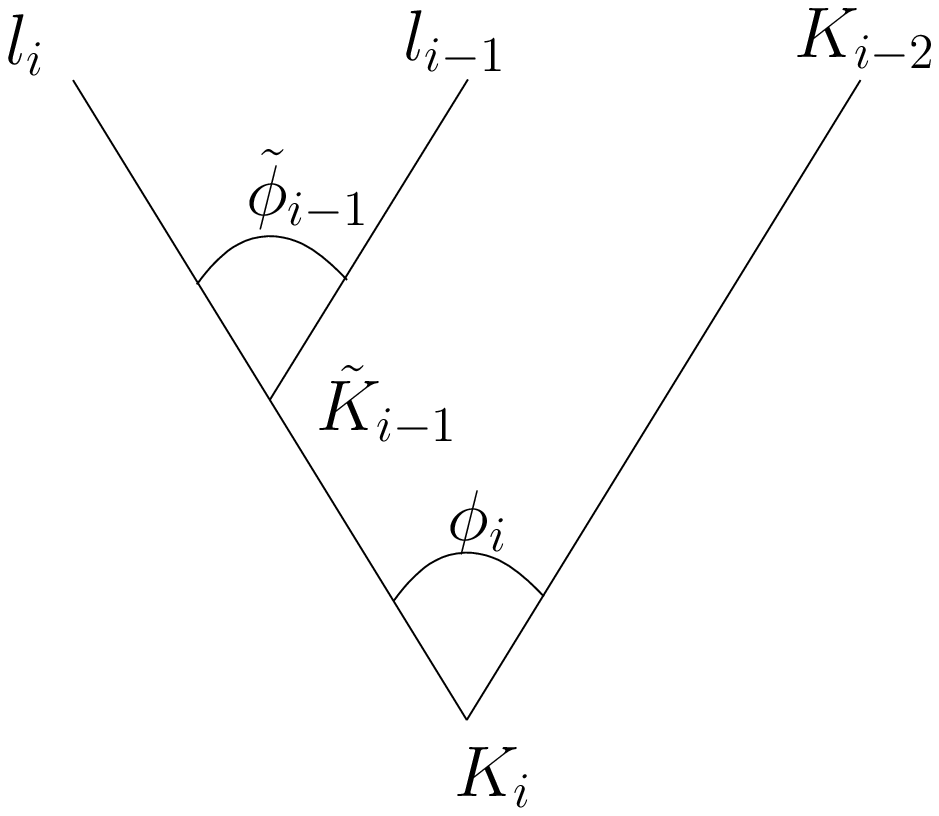} 
   \end{minipage} \,.
\end{equation}
Here $K_i=K_{i-1}+l_i+2n_i=\tilde K_{i-1}+l_i+2\tilde n_i$
The explicit definition of the coefficients is given in Ref.~\cite{gattobigio_nonsymmetrized_2011}.
Let us call ${\mathcal Y}^{LM}_{[K]}(\Omega^i_N)$ the HH basis element
constructed in terms of a set of Jacobi coordinates in which the $i$-th
and $i+1$-th Jacobi vectors results from the 
transposition between particles $j,j+1$
\begin{equation}
  \begin{aligned}
 \mathbf x'_{i}  &= - \frac{1}{j} \,\mathbf x_i + 
                      \frac{\sqrt{(j+1)^2-2(j+1)}}{j}  \,\mathbf x_{i+1} \\
 \mathbf x'_{i+1}&=   \frac{\sqrt{(j+1)^2-2(j+1)}}{j} \,\mathbf x_i 
                    + \frac{1}{j} \,\mathbf x_{i+1}   \,,
  \end{aligned}
  \label{eq:jc3}
\end{equation}
with all the
other vectors equal to the original ones (transposed basis). 
The coefficients
\begin{equation}
 {\mathcal A}^{i,LM}_{[K][K']}=\int d\Omega_N[{\mathcal Y}^{LM}_{[K]} (\Omega_N)]^*
 {\mathcal Y}^{LM}_{[K']}(\Omega^i_N)\,,
\label{eq:ca1}
\end{equation}
are the matrix elements of a matrix ${\mathcal A}^{LM}_i$
 that allows to express the transposed HH basis 
elements in terms of the reference basis. 
The coefficients ${\mathcal A}^{i,LM}_{[K][K']}$ form a very sparse matrix and
they can be calculated analytically using the
${\cal T}$- coupling coefficients
and the Raynal-Revai matrix elements~\cite{gattobigio_nonsymmetrized_2011} .
A generic rotation between the
reference HH basis and a basis in which the last
Jacobi vector is defined as $\mathbf x'_N=\mathbf r_j-\mathbf r_i$
can be constructed as successive products of the 
${\mathcal A}^{k,LM}_{[K][K']}$ coefficients. 
Defining ${\mathcal Y}^{LM}_{[K]}(\Omega^{ij}_N)$ the HH basis element
constructed in terms of a set of Jacobi coordinates in which the 
$N$-th Jacobi vector is defined $\mathbf x'_N=\mathbf r_j-\mathbf r_i$,
this coefficient can be given in the following form
\begin{equation}
 {\mathcal B}^{ij,LM}_{[K][K']}=\int d\Omega[{\mathcal Y}^{LM}_{[K]} (\Omega_N)]^*
 {\mathcal Y}^{LM}_{[K]}(\Omega^{ij}_N) =
\left[{\mathcal A}^{LM}_{i_1}\cdots{\mathcal A}^{LM}_{i_n}\right]_{[K][K']} \,.
\label{eq:ca2}
\end{equation}
The particular values of the indices $i_1,\ldots,i_n$, labelling
the matrices ${\mathcal A}^{LM}_{i_1},\ldots,{\mathcal A}^{LM}_{i_n}$, 
depend on the pair $(i,j)$.
The matrix
\begin{equation}
{\mathcal B}_{ij}^{LM}={\mathcal A}^{LM}_{i_1}\cdots{\mathcal A}^{LM}_{i_n}\,,
\label{eq:matrixb}
\end{equation}
is written as a product of the sparse matrices ${\mathcal A}^{LM}_{i}$'s. 

We consider now the potential energy of an $A$-body system constructed in terms 
of two-body interactions
\begin{equation}
   V=\sum_{i<j} V(i,j)  \;\;\; .
\end{equation}
In terms of the HH basis it results
\begin{equation}
\sum_{ij} V_{ij}(\rho)=\sum_{ij} 
[{\cal B}^{LM}_{ij}]^t\, V_{12}(\rho)\,{\cal B}^{LM}_{ij} \,.
\label{eq:vpot}
\end{equation}
where the matrix elements of the matrix $V_{12}(\rho)$ are defined as
\begin{equation}
\begin{aligned}
& V^{(1,2)}_{[K][K']}(\rho)=
\langle{\cal Y}^{LM}_{[K]}(\Omega_N)|V(1,2)|{\cal Y}^{LM}_{[K']}(\Omega_N)\rangle= \cr
&\delta_{l_1,l^\prime_1}\cdots\delta_{l_N,l^\prime_N}
\delta_{L_2,L^\prime_2}\cdots\delta_{L_N,L^\prime_N}
\delta_{K_2,K^\prime_2}\cdots\delta_{K_N,K^\prime_N} \cr
&\times \int d\phi_N(\cos\phi_N\sin\phi_N)^2
\;{\cal P}^{\alpha_{l_N},\alpha_{K_{N-1}}}_{K_N}(\phi_N)
\,V(\rho\cos\phi_N)\,{\cal P}^{\alpha_{l_N},\alpha_{K_{N-1}}}_{K'_N}(\phi_N)\,.
\end{aligned}
\label{eq:v12}
\end{equation}

Each term of the sum in Eq.(\ref{eq:vpot}) results in a product of sparse
matrices, a property which allows an efficient implementation of matrix-vector
product. This procedure can be easily extended to spin-dependent
potentials~\cite{gattobigio_non-symmetrized_2009}.

We now consider a three body force depending on the hyperradius
$\rho_{ijk}$ of a triplet of particles $\mathbf r_i,\mathbf r_j,\mathbf r_k$,
\begin{equation}
  V^{(3)} = \sum_{i<j<k} W(\rho_{ijk})\,.
  \label{}
\end{equation}
The term in which $i,j,k\equiv 1,2,3$ verifies $\rho^2_{123} = x^2_N+x^2_{N-1}$.
It can be calculated on a hyperspherical-basis set relative to an
non-standard hyperspherical tree with the branches attached to leaves $x_N$ and
$x_{N-1}$ going to the same node. The
transition between the two hyperspherical sets is given by 
the ${\cal T}$-coefficients
\begin{equation}
  {\cal Y}^{LM}_{[K]}(\Omega_N)
   =
   \sum_{\tilde n_{N-1}} 
   {\cal T}^{\alpha_{K_{N-2}}\alpha_{l_{N-1}}\alpha_{l_N}}_{n_{N-1} \;\tilde
   n_{N-1}\; K}
   {\cal Y}^{LM}_{[\tilde K]}(\tilde\Omega_N) \,,
  \label{}
\end{equation}
where all the variable with the tilde refer to the non-standard tree.
With this choice we simply have
\begin{equation}
  \rho_{123} = \rho\cos\phi_N\,,
  \label{}
\end{equation}
and the fixed-rho matrix elements reads
\begin{equation}
  \begin{aligned}
    \langle {\cal Y}^{LM}_{[\tilde K']}(\tilde\Omega_N) | W(\rho) | {\cal Y}^{LM}_{[\tilde K]}(\tilde\Omega_N) \rangle
  &= 
  \delta_{l'_1,l_1} \cdots \delta_{l'_N,l_N}
  \delta_{L'_2,L_2} \cdots \delta_{L',L} \delta_{M',M}
  \delta_{\tilde K'_2,\tilde K_2} \cdots \delta_{\tilde K'_{N-1},\tilde K_{N-1}}  \\
  \int (\cos\phi_N)^{4}(\sin\phi_N)^{3N-8} d\phi_N\;
  &
  {\mathcal P}_{K'}^{\alpha_{\tilde K_{N-1}},\alpha_{K_{N-2}}}(\phi_N)  
  {\mathcal P}_{K}^{\alpha_{\tilde K_{N-1}},\alpha_{K_{N-2}}}(\phi_N) 
  W(\rho\cos\phi_N) \,.
  \label{}
  \end{aligned}
\end{equation}
The three-body force matrix $W(\rho)$ is extremely
sparse, and it is diagonal on all quantum numbers but the grand-angular
momentum. Finally the matrix $W(\rho)$ in the standard basis is obtained by means of 
the ${\cal T}$-coefficients
\begin{equation}
  \begin{aligned}
    \langle {\cal Y}^{LM}_{[K']}(\Omega_N) | W(\rho) | {\cal Y}^{LM}_{[K]}(\Omega_N) \rangle
  &= 
   \sum_{\tilde n_{N-1}} 
   {\cal T}^{\alpha_{K_{N-2}}\alpha_{l_{N-1}}\alpha_{l_N}}_{n'_{N-1} \;\tilde
   n_{N-1}\; K'}
   {\cal T}^{\alpha_{K_{N-2}}\alpha_{l_{N-1}}\alpha_{l_N}}_{n_{N-1} \;\tilde
   n_{N-1}\; K}
    \langle {\cal Y}^{LM}_{[\tilde K']}(\tilde\Omega_N) | W(\rho_{123}) | {\cal
    Y}^{LM}_{[\tilde K]}(\tilde\Omega_N) \rangle \,.
  \end{aligned}
\end{equation}

In order to calculate the other terms of the three-body force, we use the 
matrices ${\cal A}_p^{LM}$, defined in Eq.~(\ref{eq:ca1}), that transpose
particles; with a suitable product of these sparse matrices
\begin{equation}
  {\cal D}_{ijk}^{LM} =  {\cal A}_{p_1}^{LM} \cdots {\cal A}_{p_m}^{LM}\,,
  \label{}
\end{equation}
we can permute the
particles in such a way that $\mathbf x_N = \mathbf r_i-\mathbf r_j$, 
and $\mathbf x_{N-1} = 2/\sqrt{3}(\mathbf r_k - (\mathbf r_i+\mathbf r_j)/2)$,
and $\rho^2_{ijk} = x_{N-1}^2 +  x_N^2$, and the total three-body force reads
\begin{equation}
  V^{(3)} = \sum_{i<j<k} [{\cal D}_{ijk}^{LM}]^t\, W(\rho) \,
  {\cal D}_{ijk}^{LM}  \,.
  \label{}
\end{equation}

\newpage

\newpage

\begin{table}
  \caption{The ground state $E^{(0)}_{3b}$ and the excited state $E^{(1)}_{3b}$
of the helium trimer calculated with the LM2M2 potential,
with its gaussian representation and with
the gaussian potential plus the three-body potential. In this case
the two parameters, the strength $W_0$ and the range $\rho_0$ are given.}
\label{tab:table2}
\begin{center}
\begin{tabular}{@{}ccc}
\hline
potential & $E^{(0)}_{3b}$ (mK) &  $E^{(1)}_{3b}$ (mK)   \cr
\hline
 LM2M2 \cite{kolganova_ultracold_2009}   & $-126.4$ & $-2.265$    \cr
 gaussian & $-150.4$ & $-2.467$   \cr
\hline
 ($W_0$ (K), $\rho_0$ (a.u.)) &         &  \cr
 $(306.9,4)$    & $-126.4$ & $-2.283$    \cr
 $(18.314,6)$   & $-126.4$ & $-2.287$    \cr
 $(4.0114,8)$   & $-126.4$ & $-2.289$    \cr
 $(1.4742,10)$  & $-126.4$ & $-2.292$    \cr
 $(0.721 ,12)$  & $-126.4$ & $-2.295$    \cr
 $(0.422 ,14)$  & $-126.4$ & $-2.299$    \cr
 $(0.279 ,16)$  & $-126.4$ & $-2.302$    \cr
\hline
\end{tabular}
\end{center}
\end{table}

\begin{table}
  \caption{$A=4,5,6$ binding energies of the ground, $E^{(0)}_{Ab}$, and the
  first two excited states, $E^{(1)}_{Ab}$ and $E^{(2)}_{Ab}$, for increasing
  values of the grand-angular-quantum number $K$ using the two-body soft-core
  gaussian potential. We also report the symmetry of the states; the ground,
  $E^{(0)}_{Ab}$, and the first-exited, $E^{(1)}_{Ab}$, states are totally symmetric; 
  the second-exited state belongs to a mixed representation.}
\label{tab:table1}
\begin{center}
\begin{tabular}{@{}c|ccc|ccc|ccc}
\hline
$K$ & $E^{(0)}_{4b}(mK)$ &  $E^{(1)}_{4b}(mK)$ & $E^{(2)}_{4b}(mK)$ 
& $E^{(0)}_{5b}(mK)$ &  $E^{(1)}_{5b}(mK)$ &   $E^{(2)}_{5b}(mK)$ 
& $E^{(0)}_{6b}(mK)$ &  $E^{(1)}_{6b}(mK)$ &  $E^{(2)}_{6b}(mK)$ \cr
& [\bf 4] & [\bf 4] & [\bf {3\;1}] 
& [\bf 5] & [\bf 5] & [\bf {4\;1}] 
& [\bf 6] & [\bf 6] & [\bf {5\;1}] \\
\hline
0      & 725.98 & 31.688 &        & 1913.0 & 642.84 &        & 3773.1 & 2010.7 &       \cr
2      & 725.98 & 31.688 &        & 1913.0 & 642.84 & 314.15 & 3773.1 & 2010.9 & 1626.5\cr
4      & 746.45 & 77.971 &        & 1941.2 & 746.01 & 400.95 & 3807.6 & 2140.1 & 1719.3\cr
6      & 750.15 & 107.63 &        & 1944.1 & 778.79 & 516.60 & 3809.9 & 2166.2 & 1840.5\cr
8      & 751.06 & 124.48 & 2.5177 & 1945.0 & 802.47 & 571.03 & 3810.8 & 2188.6 & 1882.5\cr
10     & 751.28 & 135.94 & 29.401 & 1945.2 & 813.88 & 608.58 & 3810.9 & 2196.4 & 1909.0\cr
12     & 751.35 & 144.17 & 50.336 & 1945.2 & 820.87 & 634.25 & 3810.9 & 2200.8 & 1923.4\cr
14     & 751.37 & 149.30 & 66.672 & 1945.2 & 824.84 & 653.19 & 3810.9 & 2202.7 & 1931.9\cr
16     & 751.37 & 152.98 & 79.082 & 1945.2 & 827.23 & 657.59 & 3810.9 & 2203.6 & 1936.5\cr
18     & 751.38 & 155.54 & 89.069 & 1945.2 & 828.67 & 678.86 & 3810.9 & 2204.0 & 1938.7\cr
20     & 751.38 & 157.43 & 97.021 & 1945.2 & 829.58 & 687.87 & 3810.9 & 2204.1 & 1939.7\cr
22     & 751.38 & 158.76 & 103.54 & 1945.2 & 830.15 & 695.23 & 3810.9 & 2204.2 & 1940.0\cr
24     & 751.38 & 159.77 & 108.90 & 1945.2 & 830.50 & 701.29 &        &        &       \cr
26     & 751.38 & 160.53 & 113.38 &        &        &        &        &        &       \cr
28     & 751.38 & 161.10 & 117.16 &        &        &        &        &        &       \cr
30     & 751.38 & 161.54 & 120.37 &        &        &        &        &        &       \cr
32     & 751.38 & 161.89 & 123.13 &        &        &        &        &        &       \cr
34     & 751.38 & 162.15 & 125.52 &        &        &        &        &        &       \cr
36     & 751.38 & 162.37 & 127.60 &        &        &        &        &        &       \cr
38     & 751.38 & 162.53 & 129.42 &        &        &        &        &        &       \cr
40     & 751.38 & 162.67 & 131.02 &        &        &        &        &        &       \cr
\hline
\end{tabular}
\end{center}
\end{table}

\begin{table}
  \caption{$A=4,5,6$ binding energies of the ground, $E^{(0)}_{Ab}$, and
  first-excited, $E^{(1)}_{Ab}$, states for increasing values of the
  grand-angular-quantum number $K$. The three-body force parameters are
  $\rho_0=14$~a.u. and $W_0=0.422$~K. We also report the symmetry of the
  states; both the ground, $E^{(0)}_{Ab}$, and the first-exited,
  $E^{(1)}_{Ab}$, states are totally symmetric.}
\label{tab:table3}
\begin{center}
\begin{tabular}{@{}c|cc|cc|cc}
\hline
$K$ & $E^{(0)}_{4b}$(mK) & $E^{(1)}_{4b}$(mK) & $E^{(0)}_{5b}$(mK) 
& $E^{(1)}_{5b}$(mK) & $E^{(0)}_{6b}$(mK) & $E^{(1)}_{6b}$(mK) \cr
& [\bf 4] & [\bf 4]  
& [\bf 5] & [\bf 5]  
& [\bf 6] & [\bf 6]  \\
\hline
0      & 538.93 & 4.557  & 1288.1 & 365.1  & 2293.8 & 1109.9   \cr
2      & 538.93 & 4.557  & 1288.1 & 365.1  & 2293.8 & 1109.9   \cr
4      & 561.69 & 40.29  & 1319.6 & 460.4  & 2331.8 & 1237.3   \cr
6      & 566.68 & 67.47  & 1324.4 & 497.6  & 2336.6 & 1273.0   \cr
8      & 568.21 & 84.22  & 1326.1 & 527.0  & 2338.4 & 1307.7   \cr
10     & 568.58 & 96.04  & 1326.5 & 542.7  & 2338.7 & 1323.1   \cr
12     & 568.73 & 105.30 & 1326.6 & 554.0  & 2338.8 & 1334.4   \cr
14     & 568.77 & 111.17 & 1326.6 & 561.0  & 2338.9 & 1340.9 \cr
16     & 568.78 & 115.58 & 1326.6 & 565.9  & 2338.9 & 1345.3 \cr
18     & 568.79 & 118.78 & 1326.6 & 569.3  & 2338.9 & 1348.2 \cr
20     & 568.79 & 121.20 & 1326.6 & 571.8  & 2338.9 & 1350.2 \cr
22     & 568.79 & 122.98 & 1326.6 & 573.6  & 2338.9 & 1351.6 \cr
24     & 568.79 & 124.38 & 1326.6 & 574.9  &        &        \cr
26     & 568.79 & 125.47 &        &        &        &       \cr
28     & 568.79 & 126.33 &        &        &        &       \cr
30     & 568.79 & 127.02 &        &        &        &       \cr
32     & 568.79 & 127.57 &        &        &        &       \cr
34     & 568.79 & 128.02 &        &        &        &       \cr
36     & 568.79 & 128.40 &        &        &        &       \cr
38     & 568.79 & 128.70 &        &        &        &       \cr
40     & 568.79 & 128.96 &        &        &        &       \cr
\hline
Ref.\cite{lewerenz_structure_1997}    & 558.4  &        & 1302.2 &        & 2319.4  &        \cr
Ref.\cite{blume_monte_2000}& 559.7  & 132.6  & 1309.3 & 597.1  & 2329.4  & 1346.7 \cr
\hline
\end{tabular}
\end{center}
\end{table}

\begin{table}
  \caption{The ratios $E^{(0)}_{Ab}/E^{(0)}_{3b}$ and $E^{(1)}_{Ab}/E^{(0)}_{(A-1)b}$ as a function
 of the three-body cutoff $\rho_0$.}
\label{tab:table4}
\begin{center}
\begin{tabular}{@{}ccccccc}
\hline
$\rho_0$ (a.u.) & $E^{(0)}_{4b}/E^{(0)}_{3b}$ & $E^{(1)}_{4b}/E^{(0)}_{3b}$ 
& $E^{(0)}_{5b}/E^{(0)}_{3b}$ & $E^{(1)}_{5b}/E^{(0)}_{4b}$ 
& $E^{(0)}_{6b}/E^{(0)}_{3b}$ & $E^{(1)}_{6b}/E^{(0)}_{5b}$  \cr
\hline
  $12$            & $4.47$ & $1.01$ & $10.33$ & $1.001$ & $18.12$ & $1.005$  \cr   
  $14$            & $4.50$ & $1.02$ & $10.50$ & $1.011$ & $18.50$ & $1.018$  \cr   
  $16$            & $4.54$ & $1.03$ & $10.70$ & $1.021$ & $19.06$ & $1.029$  \cr   
\hline
\end{tabular}
\end{center}
\end{table}
\clearpage

\end{document}